\documentclass[runningheads]{llncs}
\usepackage[T1]{fontenc}
\usepackage{graphicx}
\usepackage{booktabs} 
\usepackage{multirow}
\usepackage{amsmath}
\usepackage{multibib}
\usepackage{algorithm}
\usepackage{algpseudocode}
\usepackage{hyperref}
\usepackage[table,xcdraw]{xcolor}
\usepackage{multirow}
\usepackage{tabularx} 
\usepackage{xcolor}
\usepackage[table]{xcolor}
\usepackage{float}
\usepackage{gensymb}
\usepackage{notoccite}

\usepackage{parskip}
\begin{document}
\title{Classification of Mitral Regurgitation from Cardiac Cine MRI using Clinically-Interpretable Morphological Features}
%
\author{Y. On \inst{1} \and 
K. Vimalesvaran \inst{1} \and
S. Zaman \inst{2} \and
M. Shun-Shin \inst{1} \and
J. Howard \inst{1} \and 
N. Linton \inst{1} \and
G. Cole \inst{2} \and 
A.A. Bharath \inst{1} \and 
M. Varela \inst{1,3}}

\authorrunning{On et al.}
\titlerunning{Classification of Mitral Regurgitation from Cine MRI}

\institute{Imperial College London, Exhibition Road, London, SW7 2AZ, UK \and Imperial College Healthcare NHS Trust, Du Cane Road, London, W12 0HS, UK \and Cardiovascular and Genomics Research Institute, City St George's University of London, SW17 0RE, UK
\newline
\email{yu.on16@imperial.ac.uk}}
\date{Aug 2024}
\maketitle              
\begin{abstract}
The assessment of mitral regurgitation (MR) using cardiac MRI, particularly Cine MRI, is a promising technique due to its wide availability. However, some of the temporal information available in clinical Cine MRI may not be fully utilised, as it requires detailed temporal analysis across different cardiac views. We propose a new approach to identify MR which automatically extracts 4-dimensional (3D + Time) morphological features from the reconstructed mitral annulus (MA) using Cine long-axis (LAX) views MRI.

Our feature extraction involves locating the MA insertion points to derive the reconstructed MA geometry and displacements, resulting in a total of 187 candidate features. We identify the 25 most relevant mitral valve features using minimum-redundancy maximum-relevance (MRMR) feature selection technique. We then apply linear discriminant analysis (LDA) and random forest (RF) model to determine the presence of MR. Both LDA and RF demonstrate good performance, with accuracies of $0.72\pm0.05$ and $0.73\pm0.09$, respectively, in a 5-fold cross-validation analysis.

This approach will be incorporated in an automatic tool to identify valvular diseases from Cine MRI by integrating both handcrafted and deep features. Our tool will facilitate the diagnosis of valvular disease from conventional cardiac MRI scans with no additional scanning or image analysis penalty.

All code is made available on an open-source basis at: \url{https://github.com/HenryOn2021/MA_Morphological_Features}.

\end{abstract}

\section{Introduction}
\subsubsection{Mitral Regurgitation}
Mitral regurgitation (MR) is one of the most common cardiac conditions: it currently affects $~31\%$ of the population in Europe~\cite{Vahanian20222021EACTS}, and its incidence is rising due to increases in life expectancy. MR occurs when the mitral valve (MV) leaflets do not close properly, leading to the reflow of blood from the left ventricle (LV) to the left atrium (LA) during systole, which can be categorised as primary or secondary. In the former case, it is usually caused by degenerative MV and patients can be asymptomatic with normal LV function~\cite{Ennis2010ChangesDisease}; whereas the latter case involves the deformation of the mitral annulus (MA) caused by LV dysfunction~\cite{ElSabbagh2018MitralDirections}.\\

According to clinical guidelines~\cite{Vahanian20222021EACTS,Robinson2021TheEchocardiography}, transthoracic echocardiography (TTE) is still used as the first choice for diagnosis of MR. However, it lacks reproducibility, subjectivity, and is highly dependent on operator skills~\cite{Enriquez-Sarano2009MitralRegurgitation,Caruthers2003PracticalEchocardiography,Marwick2015RecommendationsASE}, especially for the detection of mitral regurgitant flow~\cite{Vahanian20222021EACTS}. The assessment of MR severity is challenging even for experienced clinicians, as it comprise of qualitative, semi-quantitative, and quantitative measurements~\cite{Zoghbi2017RecommendationsResonance}. Therefore, computed tomography (CT) and cardiac magnetic resonance imaging (cMRI)~\cite{Otto20212020Disease,Garg2020AssessmentImaging} are increasingly used to characterise MR for functional analysis and evaluation of severity. There are, however, few tools to help identify MR from Cine MRI. This blocks the clinical potential of these images for MR diagnosis and management.\\

\vspace{-0.4cm}

\subsubsection{Cardiac Cine MRI}
Cine (or dynamic) MRI is one of the most commonly acquired cardiac MRI sequences, used to assess the function and morphology of the heart chambers for virtually all clinical cardiac MRI indications~\cite{Garg2020AssessmentImaging}. Conventional Cine MRI usually employs a 2D balanced steady-state free precession (bSSFP) readout reconstructed into approximately 30 or 50 cardiac phases. Cine MRI is acquired in both short-axis stacks (full coverage of the LV) and single-slice long-axis (2-, 3-, and 4-chamber) views~\cite{Sechtem1987CineFunction}.\\

\vspace{-0.2cm}

The MV leaflets are thin and protein-rich, making them hard to image using most types of MRI. Regurgitant jets can be observed as a region of reduced blood signal in bSSFP; however, they are often not clearly visible~\cite{On2023AutomaticMaps}. In contrast, the MA insertions can be reliably identified in Cine MRI~\cite{Kerfoot2021EstimationLearning,Garg2020AssessmentImaging}, opening the possibility of studying MR annulus morphology and dynamics using Cine MRI. Some past studies have studied MR patients using Cine MRI.  Xiao \textit{et al.}~\cite{Xiao2023MachineImaging} developed a semi-supervised neural network pipeline for MR classification using Cine MRI. However, despite being fully automated, it lacked clinical interpretability as we cannot explain the model's decision-making process (known as the 'black-box' problem). Wu \textit{et al.}~\cite{Wu2014EvaluationImaging} developed a semi-automatic approach to extract dynamic features from reconstructed MA using Cine MRI to identify diastolic dysfunction, which showed comparative results with TTE measurements. However, they did not include the rich morphological features from the reconstructed MA in their study. Ennis \textit{et al.}~\cite{Ennis2010ChangesDisease} used feature tracking to approximate the MA points and extract dynamic and morphological features from the reconstructed MA at end-diastole and end-systole phases. They however focused on evaluating treatment outcomes in patients with degenerative MV disease, not in MR identification. Leng \textit{et al.}~\cite{Leng2018ImagingTracking} also used feature tracking to extract both morphological and dynamics features from reconstructed MA using Cine MRI, and showed MR patients have significantly reduced mitral dynamics and mild annular deformation. However, to date, no studies propose optimally combining a spectrum of morphological features across full cardiac cycles for Cine-based MR classification.\\

\vspace{-0.4cm}

\subsubsection{Feature Selection}
Feature selection methods can be used to reduce high-dimensional data to a smaller subset of features, which can be used as inputs to different machine-learning problems. Minimal redundancy and maximal relevance (MRMR) is a filter-based feature selection method which ranks features based on two components - relevance and redundancy~\cite{Ding2003MinimumData,HanchuanPeng2005FeatureMin-redundancy}. In an iterative approach, MRMR uses F-statistics or mutual information to identify the features that correlate the most (are most relevant) with a given binary label (e.g., a clinical diagnosis)~\cite{Weir1984EstimatingStructure,Kraskov2004EstimatingInformation}. The feature is added to an optimal feature set, in which Pearson correlation~\cite{LeeRodgers1988ThirteenCoefficient} or mutual information is used to remove highly correlated features with lower relevance (to minimise redundancy).\\

\vspace{-0.4cm}

\subsubsection{Machine Learning}
We propose to use Linear Discriminant Analysis (LDA)~\cite{Mohammed2020LinearAnalysis} and Random Forest (RF)~\cite{Biau2016ATour} to evaluate the performance of the imaging features for MR detection. LDA looks for a linear decision boundary in feature space by fitting the class-conditional densities to the data using Bayes's rule~\cite{Mohammed2020LinearAnalysis}. RF is an ensemble learning algorithm that combines the outcomes of multiple decision trees to make predictions. A random subset of the features is trained in each decision tree to reduce over-fitting and improve accuracy.\\

\vspace{-0.4cm}

\subsubsection{Aim}
We propose a machine-learning approach to detect mitral regurgitation using imaging features extracted from long-axis (2-chamber, 3-chamber and 4-chamber) bSSFP Cine MRI.\\

\vspace{-0.2cm}

\section{Methods}
\subsubsection{Patient Demographic}
187 subjects referred for clinical cardiac MRI, consisting of 98 with no MR (mean age $54\pm17$ years, 43\% female) and 89 with MR ($63\pm16$ years, 38\% female), were imaged under ethical approval in this retrospective study. Clinical diagnostic reports based on TTE were used to divide patients into each cohort. Among the 89 MR cases, $73\%$ have mild severity, $17$\% moderate, and $9\%$ are severe. The patient demographics and volumetric characteristics are shown in \textbf{Supplementary Table 1}.\\

\vspace{-0.4cm}

\subsubsection{Image Acquisition}
Conventional breath-held, gated bSSFP Cine CMR~\cite{Herzog2013CardiovascularResonance} was performed using 1.5 or 3T Siemens (Erlangen, Germany) Prisma or Aera MRI scanners in three different hospitals. Long-axis views were acquired and reconstructed to 30 cardiac phases at a spatial resolution of $0.7 - 2.1 \times 0.7 - 2.1 \times 5.0~mm^3$. The cardiac phases were ordered such that phase 0 corresponds to the electrocardiography R-peak (end-diastole).\\

\vspace{-0.4cm}

\subsubsection{Data Analysis}
\begin{figure}[h!]
\includegraphics[width=\textwidth]{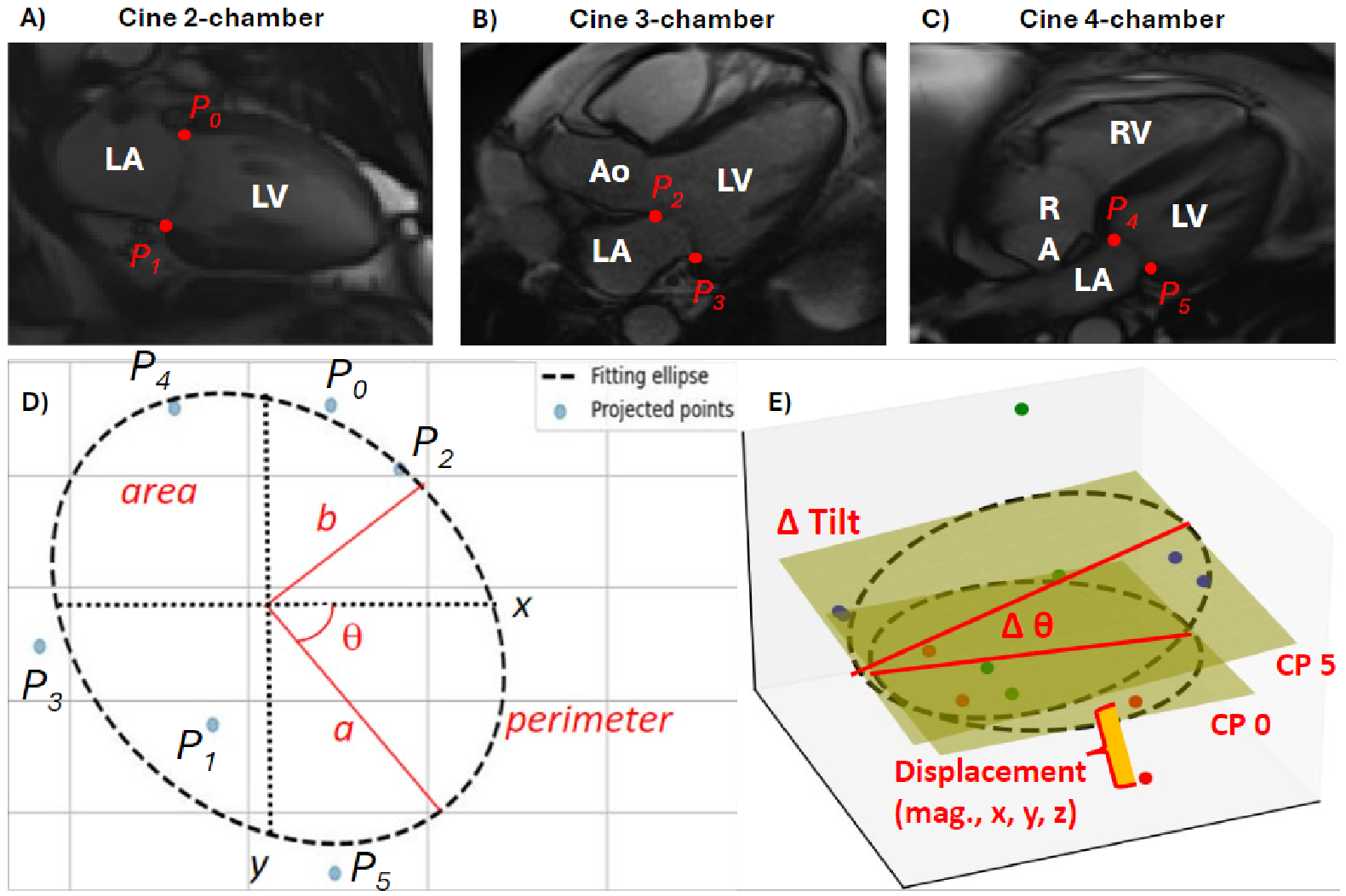}
\caption{\textbf{(A-C)} The mitral annulus (MA) insertion points are labelled (red circles) in: A) 2-chamber, B) 3-chamber and C) 4-chamber long-axis Cine views. \textbf{(D)} Example of 6 MA points in scanner coordinates (blue dots) projected to 2D space and the corresponding best-fit ellipse (black dashed line). Some of the measured ellipse properties: semi-minor axis (of length b), semi-major axis (of length a), in-plane rotation angle \(\theta\), area, and perimeter are highlighted in red. \textbf{(E)} The same set of points in 3D space, alongside the best-fit ellipse and the best-fit plane (yellow surface) estimated with SVD. Two example phases are used to illustrate some of the features being extracted in 3D space. \textit{LA: left atrium, LV: left ventricle, Ao: aorta, RA: right atrium, RV: right ventricle; $P_{0}$: mitral anterior (2-ch); $P_{1}$: mitral posterior (2-ch); $P_{2}$: mitral septal (3-ch); $P_{3}$: mitral free wall (3-ch); $P_{4}$: mitral septal (4-ch); $P_{5}$: mitral free wall (4-ch); Mag.: magnitude; CP: cardiac phase; $\Delta$: change in.}} \label{labels}
\end{figure}

Two mitral valve insertion points are manually labelled at every 5th cardiac phase (CP, out of 30) in each of the 2-, 3-, and 4-chamber long-axis views (see Fig.~\ref{labels}A-C). The inter-operator error in the positioning of mitral valve insertion points is assessed 20 randomly-selected patients. The coordinates of each of these 36 (= 6 points $\times$ 6 cardiac frames) points are converted to patient (world) coordinates using DICOM header information and used to study the morphology and dimensions of the mitral annulus (MA).\\

For each labelled CP, the 6 MA points are used to find the best-fit plane in 3D space and the best-fit ellipse on that plane (see Fig.~\ref{labels}D and E). From Fig.~\ref{labels}D, the MA points depict a close approximation of ellipse, which is proposed as a simple morphological approximation of the saddle-shaped MA. The fitted ellipses are co-registered to the common centroid at $(0,0,0)$ across patients, and the semi-major axis of the ellipse is aligned with the x-axis at $CP_{0}$ (see \textbf{Supplementary Fig. 1}). The best-fit plane is identified using singular value decomposition (SVD)~\cite{Klema1980TheApplications} and the ellipse is fitted by projecting the 3D points to the 2D plane using Rodrigues' rotation~\cite{Fraiture2009ARotation}. At each CP, we extract the following features from the fitted ellipse: area, perimeter, semi-minor axis length ($b$), semi-major axis length ($a$), eccentricity, $b:a$ ratio, MA heights (the sum of the most superior and the most inferior points perpendicular to the best-fit plane), and the (x,y,z) components of the normal plane vectors. We additionally estimate changes across consecutive cardiac phases in: the angle of the MA plane with a horizontal plane (tilt) and the in-plane rotation angle ($\theta$) of the ellipse's semi-major axis with respect to the x-axis. The magnitude and individual (x-, y-, z-) axis displacements of the 3D MA point positions across the cardiac cycle are also measured (see Fig.~\ref{labels}D, E).\\

\vspace{-0.4cm}

\subsubsection{Data Selection}
In total, 187 candidate features are extracted. The comprehensive list of features and the number of values measured from individual features are outlined in \textbf{Supplementary Table 2}. We select the most salient features by estimating individual MRMR coefficients between each feature and MR/no MR clinical labels~\cite{Ding2003MinimumData,HanchuanPeng2005FeatureMin-redundancy}. The top $K$ features with the highest relevance and lowest redundancy are selected as the input vectors for the classification stage. To decide how many features to keep for the subsequent analyses, we tested the $K$ = {5, 10, 25, 50} features and decided on $K$ = 25, which produced the best cross-validation (CV) results.\\

\subsubsection{Machine Learning Classification}
We have selected two machine learning models to classify the optimal feature set ($K$ = 25) into one of the two classes (No MR and MR): linear discriminant analysis (LDA)~\cite{Mohammed2020LinearAnalysis} and random forest (RF)~\cite{Biau2016ATour}. To optimise the RF model, we perform a grid search for the number of estimators, the number of features to consider when looking for the best split, maximum tree depth, and maximum leaf nodes using the training and validation set with the optimal feature set. The best RF configuration for the dataset is listed in \textbf{Supplementary Table 4}.\\

\subsubsection{Performance Assessment}
To mitigate the risk of overfitting and provide a reliable evaluation of the models, we perform stratified 5-fold cross-validation (CV) with a constant random seed (for reproducibility) and data shuffling~\cite{Arlot2010ASelection}. From the 187 cases, 10 cases from each class are randomly selected as a test set, which is not included in the 5-fold CV. Of the remaining 167 cases, 134 cases are used for training and 33 for validation in each fold. The test set is used for independent evaluation of the models to minimise data leakage. We assess the classification performance using the following metrics: accuracy, specificity, sensitivity, F1 score, and Area Under the Receiver Operating Characteristics Curve (ROC AUC)~\cite{M2015AEvaluations}.\\


\section{Results}
The proposed method can identify MR patients using simple MA morphological features extracted from long-axis Cine MRI. LDA and RF methods achieve a cross-validation accuracy of $0.72\pm0.05$ and $0.73\pm0.09$, respectively. The error in insertion point placements between two independent observers is $2.50\pm1.99 mm$, corresponding to $10.5\%$ of the mean ellipse’s semi-major axis length.\\


\subsection{Data Selection}
According to the MRMR analysis, 12 out of the 25 most relevant imaging features are showed in Table~\ref{feature_coef}, rank in descending order using the MRMR F-statistics relevance. From the LDA column, the ellipse perimeter at $CP_{5} \& CP_{10}$, the semi-major axis length (a) and the semi-minor axis length (b) at $CP_{10}$ have high absolute importance coefficients; whereas the magnitude displacement ($u_{mag}$) of \textit{P2} between $CP_{20-15}$, displacement of \textit{P5} along x-axis between $CP_{15-10}$, MA height, and the semi-minor axis length (b) at $CP_{5}$ have high coefficient values in the RF model. The remaining 13 of 25 features are listed in \textbf{Supplementary Table 3}.\\

\renewcommand{\arraystretch}{1.2}
\begin{table}
    \caption{12 out of the selected features ($K$ = 25) with the highest MRMR relevance (F-statistics), the corresponding LDA and RF models' coefficients, and the $mean \pm std$ values of the 2 cohorts. The 4 features with the highest absolute model coefficients are highlighted in \textbf{bold}. All distance measurements are in \textbf{mm}. $u_{x/y/z}$: displacement of x-/y-/z-coordinate; $u_{mag}$: displacement magnitude of x-,y-,z-coordinates; \textit{b}: semi-minor axis length; \textit{a}: semi-major axis length; \textit{CP}: cardiac phase.}\label{feature_coef}
    \centering
    \begin{tabular}{cccccc}
    \hline
    \toprule
    Feature & F-stat & LDA & RF & Value (No MR) & Value (MR) \\
    \hline
    $u_{mag \cdot P_{3} \cdot CP_{20-15}}$ & 19.3 & -0.32 & 0.06 & $5.73\pm3.57$ & $3.66\pm2.76$\\
    $u_{mag \cdot P_{2} \cdot CP_{20-15}}$ & 18.0 & -0.89 & \textbf{0.07} & $3.92\pm2.18$ & $2.70\pm1.67$\\
    $perimeter_{CP_{10}}$ & 16.1 & \textbf{3.58} & 0.03 & $129.4\pm20.3$ & $145.4\pm33.0$\\
    $area_{CP_{10}}$ & 15.2 & 0.34 & 0.03 & $1345.7\pm438.2$ & $1734.5\pm868.6$\\
    $height_{CP_{5}}$ & 15.1 & 0.39 & \textbf{0.08} & $5.99\pm3.59$ & $9.00\pm6.64$\\
    $perimeter_{CP_{5}}$ & 14.9 & \textbf{1.01} & 0.05 & $130.8\pm29.6$ & $151.6\pm43.0$\\
    $u_{mag \cdot P_{0} \cdot CP_{20-15}}$ & 14.9 & 0.26 & 0.05 & $5.06\pm4.01$ & $3.19\pm2.24$\\
    $u_{x \cdot P_{5} \cdot CP_{15-10}}$ & 14.2 & 0.57 & \textbf{0.07} & $-0.97\pm2.26$ & $0.52\pm3.08$\\
    $u_{mag \cdot P_{1} \cdot CP_{20-15}}$ & 14.1 & -0.07 & 0.04 & $4.93\pm3.29$ & $3.31\pm2.47$\\
    $a_{CP_{10}}$ & 13.9 & \textbf{-2.12} & 0.03 & $21.6\pm2.95$ & $24.1\pm5.91$\\
    $b_{CP_{5}}$ & 13.2 & -0.92 & \textbf{0.07} & $19.0\pm5.90$ & $22.7\pm7.63$\\
    $b_{CP_{10}}$ & 11.1 & \textbf{-1.76} & 0.03 & $19.5\pm4.25$ & $21.9\pm5.73$\\
    \bottomrule
    \end{tabular}
\end{table}

\vspace{-0.4cm}

\subsection{Performance Assessment}
Table~\ref{eval} summarises the evaluation metrics of the two models in classifying 'No MR' and 'MR' cohorts with the hand-engineered features. From the 5-fold CV and test set results, RF performs marginally better than LDA for all metrics.\\

\begin{table}
    \caption{Classification performance of LDA and RF with 5-fold CV (validation set) and test set.}\label{eval}
    \centering
    \begin{tabular}{c|c|c|c|c|c}
    \hline
    \toprule
    Model & Accuracy & Specificity & Sensitivity & F1-Score & AUC\\
    \hline
    LDA (Validation set) & $0.72\pm0.05$ & $0.77\pm0.08$ & $0.66\pm0.07$ & $0.69\pm0.06$ & $0.71\pm0.05$\\
    RF (Validation set) & $0.73\pm0.09$ & $0.77\pm0.09$ & $0.69\pm0.12$ & $0.71\pm0.10$ & $0.73\pm0.09$\\
    LDA (Test set) & 0.65 & 0.60 & 0.70 & 0.65 & 0.65\\
    RF (Test set) & 0.70 & 0.70 & 0.70 & 0.70 & 0.70\\
    \bottomrule
    \end{tabular}
\end{table}

\vspace{-0.2cm}

\begin{figure}[h!]
\includegraphics[width=\textwidth]{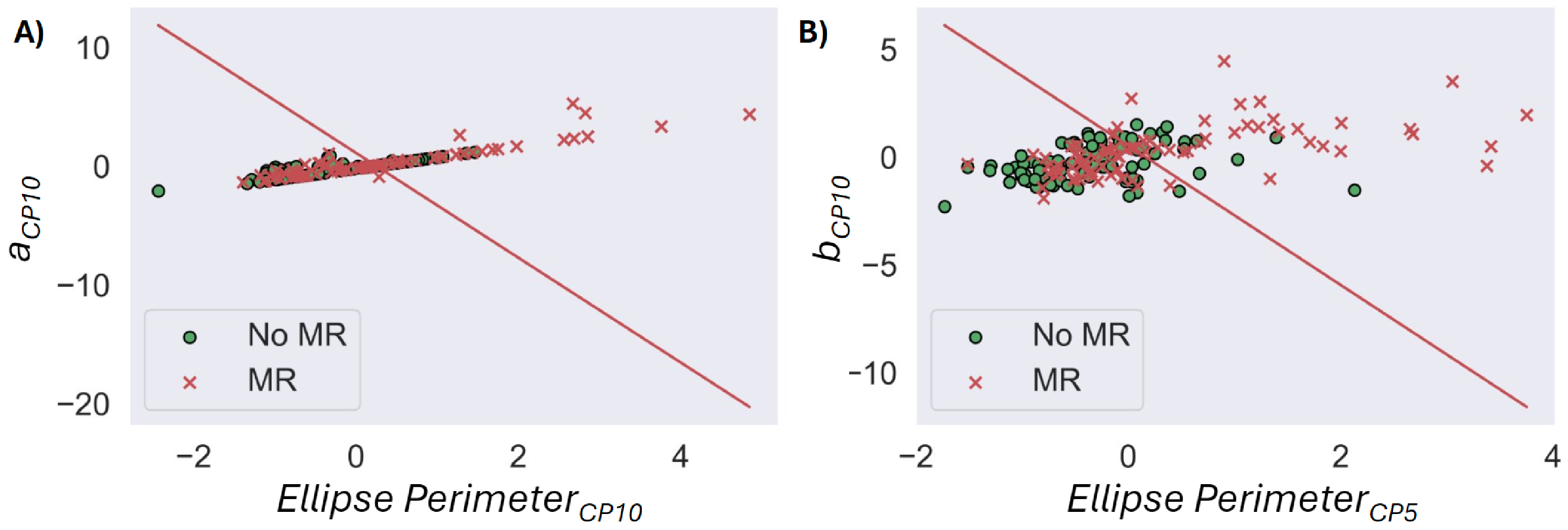}
\caption{\textbf{(A,B)} Data distribution of 'No MR' class (green circle) and 'MR' class (red cross). The hyperplane projection is calculated from the intercept and coefficients of the LDA model (fitted to the training set and predicted on the test set). The feature pairs are selected based on the LDA coefficients in descending order of absolute values.} \label{lda_plots_v2}
\end{figure}

\begin{figure}[h!]
\includegraphics[width=\textwidth]{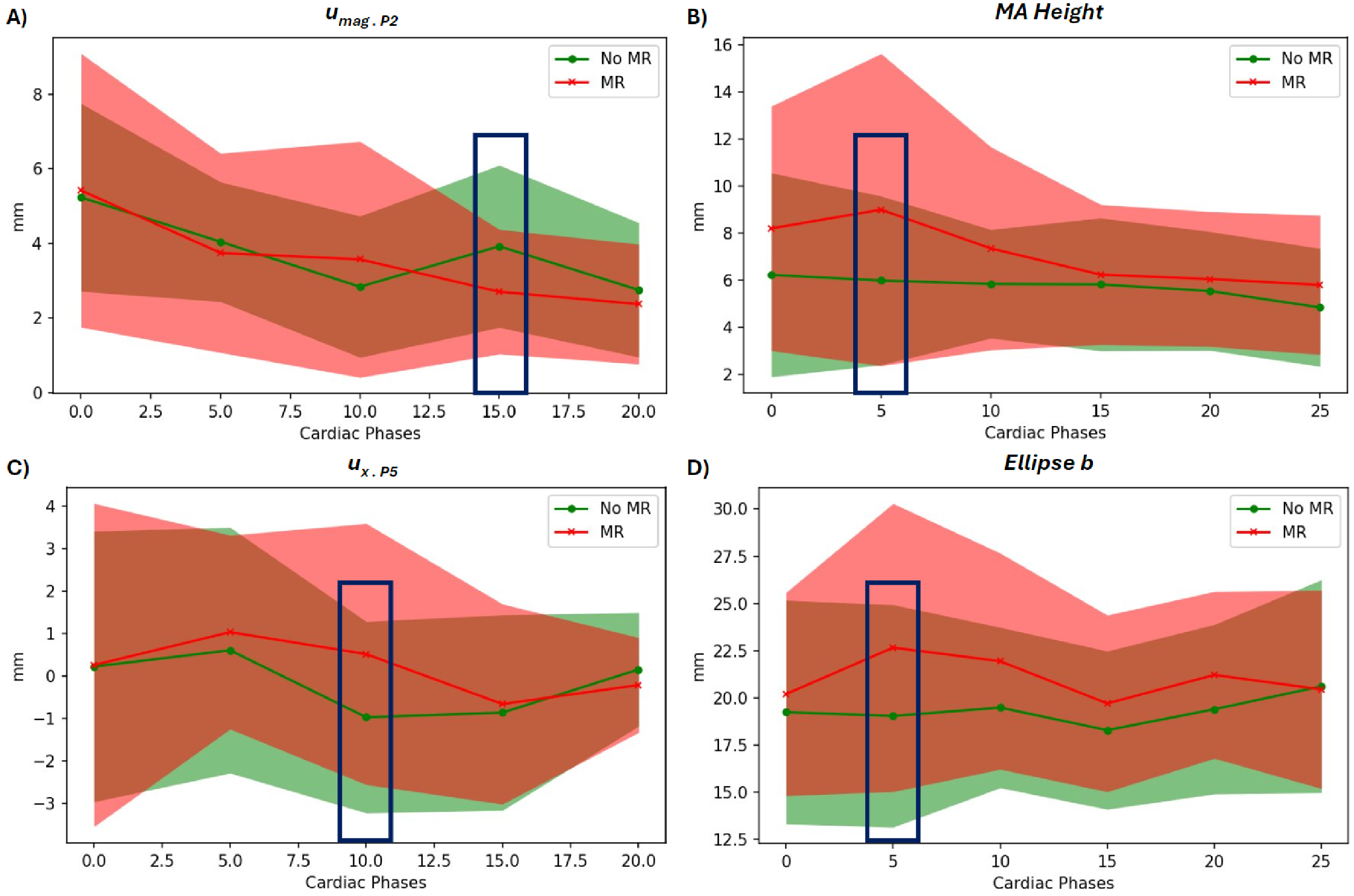}
\caption{Mean and standard deviation plots of the 4 most important RF features between the No MR (green) and MR (red) cohort. The specific CP of the selected features are highlighted in the blue box. \textbf{(A)} $u_{mag \cdot P_{2}}$ at $CP_{20-15}$; \textbf{(B)} \textit{MA Height} at $CP_{5}$; \textbf{(C)} $u_{x \cdot P_{5}}$ at $CP_{15-10}$; \textbf{(D)} $b$ at $CP_{5}$.}\label{mean_std_plots}
\end{figure}

In Fig.~\ref{lda_plots_v2}A and B, the 4 most important LDA features ($perimeter_{CP_{10}}$, $perimeter_{CP_{5}}$, $a_{CP_{10}}$, and $b_{CP_{10}}$) already show some separability between the 2 classes when considered in pairs. In Fig.~\ref{mean_std_plots}A-D, the plots illustrate the change in the 4 feature values with the highest RF coefficients across the full cardiac cycle, which further depict the class separation in the selected features at the specific CP (blue box).\\

\vspace{0.3cm}

\vspace{-0.2cm}

\section{Discussion}
In this study, we propose a semi-automatic pipeline to accurately identify mitral regurgitation from three clinical long-axis Cine MRI views. We use clinically-interpretable hand-engineered features, which can quantitatively characterise the morphology of the MA, reconstructed from 6 manually annotated MA insertions. We have not included features related to the intensity of the blood in Cine MRI, although flow-related dephasing could in principle contribute to the classification of MR~\cite{On2023AutomaticMaps}. Flow-induced dephasing of blood signal depends strongly on the imaging parameters, slice positioning and is therefore extremely variable across the patient cohort. We have performed an offline study to evaluate the flow information in classifying MR with the same cohort of patients and dataset using a 1-layer convolutional neural network (CNN). By using the labelled points to extract the flow's pixel intensities across the MV, to create a 2D array to record the intensities across the cardiac cycle. However, the CNN is unable to capture useful features to distinguish between the 2 classes, in which the performance could be hampered by the limitation of Cine MRIs, and phase contrast (i.e. pixel intensity is correlated to flow velocity) could be a better alternative MRI sequence.\\

Although the MV leaflets cannot be observed in Cine MRI, we show that these images include clinically important MA morphological information that, when optimally combined, can classify MR accurately. In Table~\ref{feature_coef}, the magnitude displacement of the labelled points in Cine 2-ch and 3-ch at mid-diastole ($CP_{20-15}$) have high relevance with the diagnostic outcomes and high importance in classification, especially in the RF model. This is illustrated in Fig.~\ref{mean_std_plots}\textbf{A}, of mitral septal in Cine 3-ch, which shows the largest class separation at the specific CP (in blue) between the 2 cohorts. The restricted motion of MA is often associated with secondary MR, which the MA becomes dilated and less dynamic~\cite{Sadeghpour2008EchocardiographicRegurgitation,Stewart1992EvaluationRegurgitation}. The semi-minor length (b) of the ellipse and mitral annulus height during systole also have high relevance and high importance in RF model. From Fig.~\ref{mean_std_plots}\textbf{B \& D}, both features show good separation between the 2 cohorts at the specific CPs. The higher feature value in \textit{b} for MR cohort reflects the dilation of the MA as it often seen in patients with MR, especially moderate to severe severity.~\cite{Silbiger2012AnatomyAnnulus}.\\


\vspace{-0.2cm}

In Fig.~\ref{lda_plots_v2}, the 2 sets of features with the highest LDA coefficients show some class separation, even without considering the other 8 features. Also, the plots show the sets of the two classes are contiguous at the LDA decision boundary. The overlapping group at the boundary are largely contributed by mild MR cases, as the MA size can still be within the normal range in primary MR patients~\cite{Robinson2021TheEchocardiography}. This poses a challenging problem to find a linear combination of features that best separates the two classes. This may suggest that the potential partitions of the feature space into conditional subgroups performed by the RF is crucial, thus explaining the better performance with the RF model.\\

\vspace{-0.2cm}

There are some limitations in the proposed pipeline. First, the landmarks are manually annotated, but we plan to make this step (and thus the entire pipeline) fully automatic. We utilise the labelled landmarks in this study as training data for a CNN that can reliably localise the landmarks from the Cine MRIs. Second, we do not assess the performance of the method separately for primary and secondary MR. We speculate that our method may be better suited for secondary MR, which is expected to bring greater morphological changes to the MA. Third, the limitations of TTE (mentioned in \textbf{Introduction}) can lead to inter- and intra-observer disagreements in MR severity stratification, and we therefore do not attempt to classify MR severity.\\

\vspace{-0.2cm}

Our future plans include using the proposed features as one of the auxiliary inputs to a deep NN pipeline that can identify multiple cardiac valvular diseases using clinical Cine MRI. This NN pipeline will potentially not only identify MR, but also classify the MR type (primary/secondary), by bringing in other types of MRI and imaging information. A more detailed automatic analysis of MA features from Cine MRI could enable accurate categorisation of the type of MR (primary/secondary) and locate the possible pathological regions. This additional knowledge is essential for disease management choices (including the types of intervention and/or medical therapy), surpassing the information that can be obtained from standalone echocardiography.

\textbf{Acknowledegments.}
This work was supported by: St George’s Hospital Charity, the NIHR Imperial Biomedical Research Centre (BRC) and the British Heart Foundation Centre of Research Excellence at Imperial College London (RE/18/4/34215). We acknowledge the computational support provided by the Imperial College Research Computing Service (DOI: 10.14469/hpc/2232).

%
%
%
\bibliographystyle{unsrt}

\bibliography{references}

\end{document}